\documentclass[aoas,MSNbibl,nameyear,seceqn,dvips]{arximspdf}
\usepackage{stfloats}
\usepackage{graphicx}

\doi{10.1214/11-AOAS500}
\volume{6}
\issue{1}
\pubyear{2012}
\firstpage{383}
\lastpage{408}

\makeatletter

\setattribute{abstract}   {width}  {297pt}

\fnbelowfloat

\newcommand{\X}{{\mathbf X}}
\newcommand{\Y}{{\mathbf Y}}
\newcommand{\A}{{\mathbf A}}
\newcommand{\B}{{\mathbf B}}

\makeatother

\begin{document}
\begin{frontmatter}

\title{Prototype selection for parameter estimation in~complex models\thanksref{T3}}
\runtitle{Prototype selection}

\thankstext{T3}{Part of this work was performed in the CDI-sponsored Center for Time Domain Informatics (\protect\url{http://cftd.info}).}

\begin{aug}
\author[A]{\fnms{Joseph W.} \snm{Richards}\corref{}\thanksref{t1}\ead[label=e1]{jwrichar@stat.berkeley.edu}},
\author[B]{\fnms{Ann B.} \snm{Lee}\thanksref{t2}},
\author[B]{\fnms{Chad~M.}~\snm{Schafer}\thanksref{t2}}
\and
\author[B]{\fnms{Peter E.} \snm{Freeman}\thanksref{t2}}
\runauthor{Richards, Lee, Schafer and Freeman}
\affiliation{University of California, Berkeley,
Carnegie~Mellon University, Carnegie~Mellon University and
Carnegie~Mellon University}
\address[A]{J. W. Richards\\
Department of Astronomy\\
University of California, Berkeley\\
601 Campbell Hall\\
Berkeley, California 94720\\
USA\\
\printead{e1}}
\address[B]{A. B. Lee\\
C. M. Schafer\\
P. E. Freeman\\
Department of Statistics\\
Carnegie Mellon University\\
5000 Forbes Avenue\\
Pittsburgh, Pennsylvania 15213\\
USA}
\end{aug}

\thankstext{t1}{Supported by a Cyber-Enabled Discovery
and Innovation (CDI) Grant 0941742 from the National Science
Foundation.}
\thankstext{t2}{Supported by ONR Grant 00424143, NSF Grant 0707059 and
NASA AISR Grant NNX09AK59G.}

\received{\smonth{7} \syear{2010}}
\revised{\smonth{5} \syear{2011}}

%
\begin{abstract}
Parameter estimation in astrophysics often requires the use of complex
physical models. In this paper we study the problem of estimating the
parameters that describe star formation history (SFH) in galaxies.
Here, high-dimensional spectral data from galaxies are appropriately
modeled as linear combinations of physical components, called simple
stellar populations (SSPs), plus some nonlinear distortions.
Theoretical data for each SSP is produced for a fixed parameter vector
via computer modeling. Though the parameters that define each SSP are
continuous, optimizing the signal model over a large set of SSPs on a
fine parameter grid is computationally infeasible and inefficient. The
goal of this study is to estimate the set of parameters that describes
the SFH of each galaxy. These target parameters, such as the average
ages and chemical compositions of the galaxy's stellar populations, are
derived from the SSP parameters and the component weights in the signal
model. Here, we introduce a principled approach of choosing a small
basis of SSP \textit{prototypes} for SFH parameter estimation. The basic
idea is to quantize the vector space and effective support of the model
components. In addition to greater computational efficiency, we
achieve better estimates of the SFH target parameters. In simulations,
our proposed quantization method obtains a substantial improvement in
estimating the target parameters over the common method of employing
a~parameter grid. Sparse coding techniques are not appropriate for this
problem without proper constraints, while constrained sparse coding
methods perform poorly for parameter estimation because their objective
is signal reconstruction, not estimation of the target parameters.
\end{abstract}

%

\end{frontmatter}

\section{Introduction}

In astronomy and cosmology one is often challenged by the complexity of
the relationship between the physical parameters to be estimated and the
distribution of the observed data.
In a typical application the mapping from the \textit{parameter space} to the
\textit{observed data space} is built on sophisticated physical theory or
simulation models or both.
These scientifically motivated models are growing ever more
complex and nuanced as a result of both increased computing power and
improved understanding of the underlying physical processes.
At the same time, data are progressively more abundant and of higher
dimensionality as a result of more sophisticated detectors and greater data
collection capacity. These challenges create opportunities for
statisticians to
make a large impact in these fields.

In this paper we address one such challenge in the field of
astrophysics. Informally, the setup can be described as follows. The
observed data vector from each source is appropriately modeled as a
constrained linear combination of a set of physical components, plus
some nonlinear distortion and noise to account for observational
effects. Call this the signal model. One also has a computer model
capable of generating a dictionary of physical components under
different settings of the physical parameters. Using this dictionary of
components, the signal model can be fitted to observed data. The
parameters of interest---which we will refer to as \textit{target}
parameters---are, however, not the parameters explicitly appearing in
the signal model, but are derived from them. The target parameters
capture the physical essence of each object under study. Our goal is to
find accurate estimates of these parameters given observed data and
theoretic models of the basic components. See (\ref{eqngenmodel}) for
the formal problem statement.

Our proposed methods choose small sets of prototypes from a large
dictionary of physical components to fit the signal model to the
observed data from each object of interest. Even though the data are
truly generated as combinations of curves from a continuous (or fine)
grid of parameters, we obtain more accurate maximum likelihood
estimates of the target parameters by using a smaller, principled
choice of prototype basis. This result is partially due to the fact
that maximum likelihood estimation (MLE) often fails when the
parameters take values in an infinite-dimensional space. In
\citet{geman1982nonparametric}, the authors suggest salvaging MLE for
continuous parameter spaces by a method of sieves
[\citet{grenander1981abstract}], where one maximizes over a
constrained subspace of the parameter space and then relaxes the
constraint as the sample size grows. Quantization is one such method
for constraining the parameter space, and the optimal number of quanta
or prototypes is then determined by the sample size; see
\citet{meinicke2002quantizing} for an example of quantized density
estimation with MLE. Our approach is based on similar ideas but our
final goal is parameter estimation rather than density estimation.
Although we do not directly tie the number of quanta to the sample
size, we do observe a similar phenomenon: In the face of limited, noisy
data, gains can be made by reducing the parameter space further prior
to finding the MLE. By deriving a small set of prototypes that
effectively cover the support of the signal model, we obtain a marked
decrease in the variance of the final parameter estimates, and only a
slight increase in bias. Furthermore, by choosing a smaller set of
prototypes, the fitting procedure becomes computationally tractable.


Our principal motivation for developing this methodology is to
understand the process of star formation in galaxies. Specifically,
researchers in this field seek to improve the physical models of galaxy
evolution so that they more accurately explain the observed patterns of
galaxy star formation history (SFH) in the Universe. The principal
idea is that each galaxy consists of a~mixture of subpopulations of
stars with different ages and compositions. By estimating the
proportion of each constituent stellar subpopulation present, we can
reconstruct the star formation rate and composition as a function of
time, throughout the life of that galaxy. This is the approach of
\textit{galaxy population synthesis} [\citet{bica1988},
\citet{pela1997}, \citet{CF2001}], whereby the observed data
from each galaxy are modeled as linear combinations of a set of
idealized simple stellar populations (SSPs, groups of stars having the
same age and composition) plus some parametrized, nonlinear
distortions. Equation (\ref{cfmodel}) shows one such galaxy population
synthesis model. The fitted parameters from this signal model allow us
to estimate the SFH target parameters of each galaxy, which are simple
functions of the parameters in this model. Astrophysicists can use
the estimated SFHs of a large sample of galaxies to better understand
the physics governing the evolution of galaxies and to constrain
cosmological models. This modeling approach has produced compelling
estimates of cosmological parameters such as the cosmic star formation
rate, the evolution of stellar mass density, and the stellar initial
mass function, which describes the initial distribution of stellar
masses in a population of stars [see \citet{asa2007} and
\citet{pant2007} for examples of such results].

SFH target parameter estimates from galaxy population synthesis are
highly dependent on the choice of SSP basis. Astronomers have the
ability to theoretically model simple stellar populations from fine
parameter grids, but much care needs to be taken to determine an
appropriate basis to achieve accurate SFH parameter estimates. In
\citet{rich2009} it was shown that better parameter estimates are
achieved by exploiting the underlying geometry of the SSP disribution
than by using SSPs from regular parameter grids. In this paper we
will further explore this problem. Our main contributions are the
following:
\begin{longlist}[(1)]
\item[(1)] to introduce prototyping as an
approach to estimating parameters derived from the signal model
parameters and to show the effectiveness of quantizing the vector space
or support of the model data,
\item[(2)] to demonstrate that sparse coding
does not work as a prototyping method without the appropriate
constraints and that constrained sparse coding methods do not perform
well for target parameter estimation, and
\item[(3)] to work out the details
of the star formation history estimation problem and obtain more
accurate estimates of SFH for galaxies than the approaches used in the
astronomy and statistics literature.
\end{longlist}

There are several other fields where observed data are commonly modeled
as linear combinations of dictionaries of theoretical or idealized
components (plus some parametrized distortions), for example:
\textit{remote sensing}, both of the Earth [\citet{robe1998}] and
other planets [\citet{adam1986}], where the observed spectrum of
each area of land is modeled as a mixture of pure spectral
``endmembers;'' \textit{computer vision and computational anatomy}
[\citet{alla2007}, \citet{sabu2008}], where data are modeled
as mixtures of deformable templates; and \textit{compositional modeling
of asteroids} [\citet{clar2004}, \citet{hapk1981}], where
observed asteroids are described as mixtures of pure minerals to
determine their composition. These applications can benefit from the
methodology proposed here. A~related and important problem in
theoretical physics is \textit{gravitational wave modeling}
[\citet{baba2006}, \citet{owen1999}], where large template
banks are used to estimate the parameters of observed compact binary
systems (such as neutron stars and black holes). In this particular
problem, one is interpolating between runs of the computer model, and
not modeling the observed data as superpositions of the model output,
as we do in this paper.


There are strong connections between this work and ongoing research into
the design of \textit{computer experiments}; see \citet{Santner2003}
and \citet{Levy2010} for an overview of the topic. The fundamental
challenge in that setting is to adequately characterize the relationship
between input parameters to a simulation model and the output that the
model produces. The term ``simulation model'' should be interpreted broadly
to mean computer code which produces output as a function of input
parameters; in situations of interest, this code is a
computationally-intensive model for a complex physical phenomenom. Hence,
one must carefully ``design the computer experiment'' by choosing the set
of input parameter vectors for which runs of the simulator will be made.
Regression methods are then used to approximate the output of the simulator
for other values of the input parameters.
As is the case in our application, the ultimate objective is to compare
observed data with the simulated output to constrain these input parameters.
Research has largely focused on situations in which the output of interest
is scalar, but there has been recent work on functional outputs; see,
for instance, \citet{Bayarri2007}. Here, we have the same goal of
parameter estimation, but instead of seeking to reduce the number of times
the computer code must be run, we instead work with the scientific details
of the problem at hand and simplify the code in a principled manner to
reduce the computational burden.

\subsection{Introductory example}
\label{secex}

\begin{figure}

\includegraphics{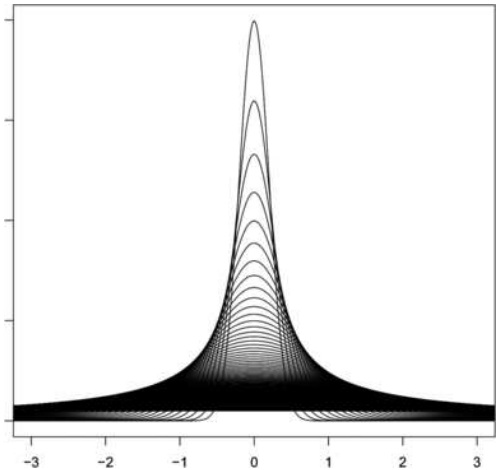}

\caption{Database of Gaussian curves used in the example in
Section \protect\ref{secex}. Simulated data are generated as noisy
random sparse
linear combinations of these curves. As $\sigma$ increases, it becomes
more difficult to distinguish the curves, especially in the presence of
noise. A basis of prototypes for estimation of the target parameter,
$\bar{\sigma}$, should include a higher proportion of low-$\sigma$
Gaussian curves.} \label{fignormcurv}
\end{figure}

To elucidate the challenges of this type of modeling problem, we begin
with a simple example. Imagine our dictionary consists of $\mu=0$
Gaussian functions generated over a fine grid of $\sigma$, such as
those in Figure~\ref{fignormcurv}. We observe a set of objects, each
producing data from a~different function constructed as a sparse linear
combination of the dictionary of Gaussian functions. The data from each
object are sampled across a fixed grid with additive i.i.d. Gaussian
noise. The component weights are constrained to be nonnegative and sum
to~1, ensuring that all parameters are physically-plausible (e.g.,
$\bar{\sigma}>0$).

Our ultimate goal is to estimate a set of target parameters for each
observed data point. In this example, our target is $\bar{\sigma}$,
the weighted average $\sigma$ of the component Gaussian curves of each
observed data vector. To this end, we model each observed curve as a
linear superposition of a set of prototypes and use the estimated
prototype weights to estimate $\bar{\sigma}$.

If our goal were to reconstruct each data point with as small of error
as possible, then a prototyping approach that samples along the
boundary of the convex hull of the dictionary of Gaussian functions
(such as archetypal analysis, see
Section~\ref{secaa}) would be optimal. In this paper, the goal is to
achieve small errors in the \textit{target parameter estimates}. A
common approach for this problem is to sample prototypes uniformly over
the parameter space. However, this often leads to the inclusion of many
prototypes with nearly identical curves. Consider the Gaussian curve
example: for high values of $\sigma$, the curves do not change
considerably with respect to changes in $\sigma$. Under the presence
of noise, curves with large $\sigma$ are not distinguishable. We are
better off including a higher proportion of prototypes in the
low-$\sigma$ range, where curves change more with respect to changes
in $\sigma$.

This intuition leads us to a different approach: choose prototypes by
quantizing the space of \textit{curves} (see
Section \ref{ssquant}). We show in Section \ref{secnorm} that a~method that selects prototypes by quantizing the vector space of
theoretical components outperforms the method of choosing prototypes
from a~uniform grid of $\sigma$ in the estimation of $\bar{\sigma}$
(see Figure \ref{figsigmse}).
Additionally, judicious selection of a reduced prototype basis
is an effective regularization of an estimation problem that is
subject to large variance when the full range of theoretical components
are utilized without any smoothing. The simulation results
shown below will display markedly reduced variances in the estimates of the
parameters of interest relative to the same procedures using larger
libraries of basis functions.

Additionally, smaller prototype bases yield better parameter estimates
than the approach of using \textit{all} of the theoretical components to
model observed data, a~phenomenon that can be explained by the markedly
reduced variance of parameter estimates found by smaller,
judiciously-chosen bases.

\subsection{Paper organization}

The paper is organized as follows. In Section~\ref{secsfh} we detail
the problem of estimating star formation history parameters for
galaxies and explain how prototyping methods can be used to obtain
accurate parameter estimates. In Section \ref{secproblem} we formalize
the problem of prototype selection for target parameter estimation and
in Section \ref{secmeth} describe several approaches. We apply those
methods to simulated data in Section \ref{secsim} to compare their
performances. In Section~\ref{secsdss} we return to the astrophysics
example, applying our methods to galaxy data from the Sloan Digital Sky
Survey. We end with some concluding remarks in Section \ref{secconc}.

\section{Modeling galaxy star formation history}
\label{secsfh}

Galaxies are gravitationally-bound objects containing $10^5$--$10^{10}$
stars, gas, dust and dark matter. The characteristics of the light we
detect from each galaxy primarily depend on the physical parameters
(e.g., age and composition) of its component stars as well as
distortions due to dust that resides in our line of sight to that
galaxy, spectral distortions due to the line-of-sight component of the
orbital velocities of its component stars, and the distance to the galaxy.

The physical mechanisms that govern galaxy formation and evolution are
complicated and poorly understood. Galaxies are complex, dynamic
objects. The star formation rate (SFR) of each galaxy tends to change
considerably throughout its lifetime and the patterns of SFR vary
greatly between different galaxies. The SFR for each galaxy depends on
a countless number of factors, such as merger history, the galaxy's
local environment (e.g., the matter density of its neighborhood, and
the properties of surrounding galaxies) and chemical composition.
Astronomers are interested in refining galaxy evolution models so that
they match the observed patterns of galaxy SFH in the Universe. It is
imperative that we first have accurate estimates of the star formation
history parameters for each observed galaxy. These SFH estimates are
necessary to test competing physical models, alert to possible
shortcomings in current models, and estimate cosmological parameters
[for an example of such an analysis, see \citet{asa2007}].

\subsection{Population synthesis model}

A common technique in the astronomy literature, called empirical
population synthesis, is to model each galaxy as a~mixture of stars
from different simple stellar populations (SSPs), defined as groups of
stars with the same age and metallicity ($Z$, defined as the fraction
of mass contributed by any element heavier than helium). The principle
behind this method is that each galaxy consists of multiple
subpopulations of stars of different age and composition so that the
integrated observed light from each galaxy is a mixture of the light
contributed by each SSP.
Describing the data from each galaxy as a combination of SSPs allows us
to reconstruct the star formation and metallicity history of each
galaxy. This is because, for each galaxy, the component weight on an
SSP captures the proportion of that galaxy's stars that was created at
the specific epoch corresponding to the age of that SSP. Therefore, the
full vector of SSP component weights for each galaxy describes the star
formation throughout the galaxy's lifetime.

\begin{figure}

\includegraphics{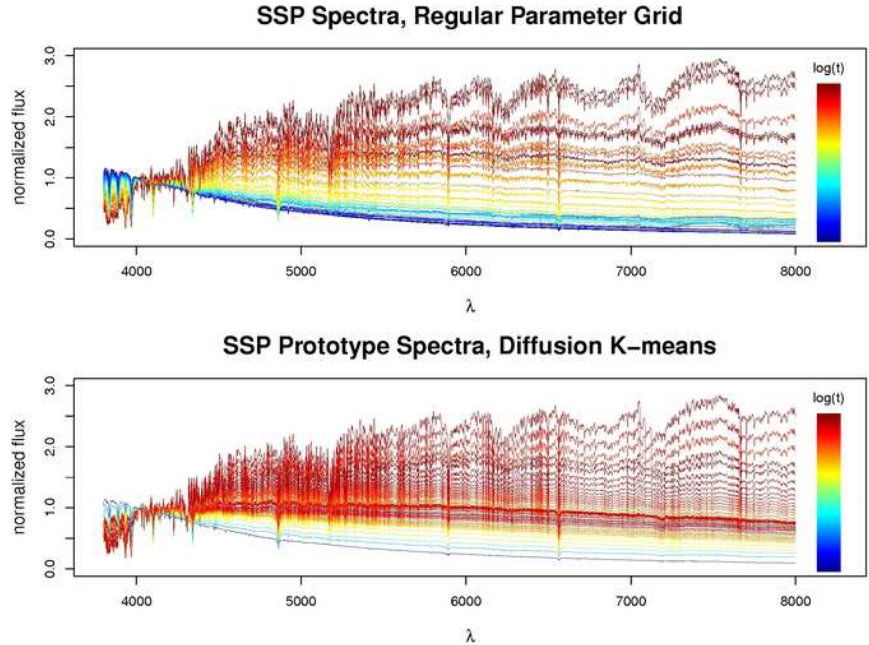}

\caption{Two bases of SSP spectra of size $K=45$, colored by $\log t$.
Each spectrum is normalized to 1 at $\lambda_0 = 4020$ \AA.
Top: basis of regular ($t,Z$) grid used in Cid Fernandes et al.
(\protect\citeyear{CF2005}). Bottom: diffusion $K$-means basis used in Richards
et al. (\protect\citeyear{rich2009}). The diffusion $K$-means basis shows a
more gradual sampling of spectral space than the regular grid basis,
which over-samples spectra from young stellar populations.}
\label{figssps}
\end{figure}

Theoretical SSPs can be produced by physical models, that are in turn
constrained by observational studies. These models typically start with
a~set of initial conditions and evolve the system forward in time based
on sets of physically motivated differential equations. The output
produced by these models can be extremely detailed. In our study, we
use a set of high-resolution, broad-band spectra from the SSP models of
\citet{BC2003}. 
See Figure \ref{figssps} for an example of some SSP spectra, plotted
over the optical portion of the electromagnetic spectrum.

The galaxy data we use to estimate SFH parameters are high-resolution,
broad-band spectra from the Sloan Digital Sky Survey [SDSS,
\citet{york2000}] which consist of light flux measurements over
thousands of wavelength bins. To model the data from each galaxy, we
adopt the empirical population synthesis generative model of a galaxy
spectrum introduced in \citet{CF2004}:
%
\begin{equation}
\label{cfmodel}\qquad
\Y_{\lambda} (\bolds{\gamma}, M_{\lambda_0}, A_V, v_*, \sigma_*)=
M_{\lambda_0}\Biggl(\sum_{j=1}^{N}\gamma_j \X_{j,\lambda}
r_{\lambda}(A_V)\Biggr) \otimes G(v_*,\sigma_*),
\end{equation}
where $\Y_{\lambda}$ is the light flux at wavelength $\lambda$.
The components of model~(\ref{cfmodel}) are the following:
\begin{itemize}
\item$\X_{j}$ is the $j$th SSP spectrum normalized at wavelength
$\lambda_0$. Each SSP has age $t(\X_j)$ and metallicity $Z(\X_j)$.
In the true generative model, $\X$ contains an infinite number of SSP
spectra over the continuous parameters of age and metallicity.
%
\item$\gamma_j \in[0,1]$, the component proportion of the $j$th SSP.
The vector $\gamma$ is the \textit{population vector} of the galaxy, the
principal parameter of interest for calculating derived parameters
describing the SFH of a galaxy.
\item$M_{\lambda_0}$, the observed
flux at wavelength $\lambda_0$.
\item$r_{\lambda}(A_V)$ accounts for
the wavelength-dependent fraction of light that is either absorbed or
scattered out of the line of sight by foreground dust. $A_V$
parametrizes the amount of this dust extinction that occurs. We adopt
the reddening model of \citet{card1989}.
\item Convolution, in
wavelength, by the Gaussian kernel $G(v_*,\sigma_*)$ describes spectral
distortions from Doppler shifts caused by the movement of stars within
the observed galaxy with respect to our line-of-sight, and is
parame\-trized by a central velocity $v_*$ and dispersion $\sigma_*$.
Previous to the analysis, care was taken to properly resample all
spectra---both the observed and model spectra---to 1 measurement per
\AA ngstr\"{o}m.\setcounter{footnote}{3}\footnote{Note that the model SSP spectra are computed
over a broader wavelength range than the observed spectra to provide an
essential wavelength cushion for the convolution.} This was done to
ensure the reliability of the spectral errors when used by the
\texttt{STARLIGHT} spectral fitting software. More details are available at
\texttt{\href{http://www.starlight.ufsc.br/papers/Manual\_StCv04.pdf}{http://www.starlight.}
\href{http://www.starlight.ufsc.br/papers/Manual\_StCv04.pdf}{ufsc.br/papers/Manual\_StCv04.pdf}}.
\end{itemize}

\subsection{SSP basis selection and SFH parameter estimation}

For each galaxy, we observe a flux, $\mathbf{O}_{\lambda}$, at each
spectral wavelength, $\lambda$, with corresponding standard error,
$\widehat{\sigma}_{\lambda}$, estimated from photon counting
statistics and characteristics of the telescope and detector.
To estimate the target SFH parameters for each galaxy, we use the
\texttt{STARLIGHT}\footnote{\texttt{STARLIGHT} can be downloaded at \url
{http://www.starlight.ufsc.br/}.}
software of \citet{CF2005}, fitting model (\ref{cfmodel}) using
maximum likelihood. The code uses a Metropolis algorithm
with simulated annealing to minimize
%
\begin{equation}
\label{chisq}
\chi^2(\bolds{\gamma},M_{\lambda_0},A_V,v_*,\sigma_*) = \sum
_{\lambda=1}^{N_{\lambda}}\biggl(\frac{\mathbf{O}_{\lambda}-\Y
_{\lambda}}{\widehat{\sigma}_{\lambda}}\biggr)^2,
\end{equation}
where $\Y_{\lambda}$ is the model flux in (\ref{cfmodel}). The
optimization routine searches for the maximum likelihood solution for
the model $\mathbf{O}_{\lambda} \sim N(\Y_{\lambda},\widehat
{\sigma}_{\lambda})$, i.i.d. for each $\lambda$. The minimization of
(\ref{chisq}) is
performed over $N+4$ parameters: $\gamma_1,\ldots,\gamma_N, M_{\lambda
_0}, A_V$, $v_*$,
and $\sigma_*$.
The speed of the algorithm scales as $\mathcal{O}(N^2)$, so it is
imperative to pick a SSP basis with a small number of spectra.

In practice, we use a basis of $K \ll N$ \textit{prototype} SSP spectra,
$\bolds{\Psi} =\{ \bolds{\Psi}_1,\ldots,\allowbreak\bolds{\Psi}_K\}$---which
can be a carefully chosen subset or a nontrivial combination of the $\X
_j$'s---and model each galaxy spectrum as
%
\begin{equation}
\label{cfmodel1}\qquad
\Y_{\lambda} (\bolds{\beta}, M_{\lambda_0}, A_V, v_*, \sigma_*)=
M_{\lambda_0}\Biggl(\sum_{k=1}^{K}\beta_k \bolds{\Psi}_{k,\lambda
} r_{\lambda}(A_V)\Biggr) \otimes G(v_*,\sigma_*),
\end{equation}
where each prototype, $\bolds{\Psi}_j$, has age $t(\bolds{\Psi
}_k)$ and metallicity $Z(\bolds{\Psi}_k)$, and\break $\sum_{k=1}^K \beta
_k = 1$.

Our goal in this analysis is to choose a suitable SSP basis to estimate
a~set of physical parameters for each galaxy. Some of the commonly-used
SFH parameters are as follows:
\begin{itemize}
\item$\langle\log t \rangle_L = \sum_{i=1}^N \gamma_i \log t(\X
_i)$, the luminosity-weighted average log age of the stars in the galaxy,
\item$ \log\langle Z \rangle_L = \log\sum_{i=1}^N \gamma_i Z(\X
_i)$, the log luminosity-weighted average metallicity of the stars in
the galaxy,
\item$\gamma_c$, a time-binned version of the population vector,
$\gamma$, and\vadjust{\goodbreak}
\item$\langle\log t \rangle_M, \log\langle Z \rangle_M$,
mass-weighted versions of the average age and metallicity of the stars
in the galaxy.
\end{itemize}
We estimate each of these parameters using the maximum likelihood
parameters from model (\ref{cfmodel1}). In \citet{rich2009}, we
introduced a method of choosing a SSP prototype basis and compared it
to bases of regular $(t,Z)$ grids that were used in previous analyses.
See Figure \ref{figssps} for a plot of two such SSP spectral bases.

\section{Formal problem statement}
\label{secproblem}

We begin with a large, fixed set of $N$ theoretical components, each
with known
parameters $\bolds{\pi}_i$ (these are the physical
properties of each component). We refer to this set as the model data.
These data can be thought of as a sample from some distribution $P_X$
in $\mathbb{R}^p$.
The model data are stored in an $p$ by $N$ matrix $\X= [\X_1,\ldots,\X
_N]$, where $p$ is the total wavelength range of the SSP spectra. We
assume that each observed data point $\Y_j$, $j=1,\ldots,M$, is generated
from the linearly separable nonlinear model
%
\begin{equation}
\label{eqngenmodel}
\Y_{j} = f\Biggl(\sum_{i=1}^N \gamma_{ij} \X_{i} ;
\bolds{\theta}_j\Biggr) + \bolds{\varepsilon}_{j},
\end{equation}
where, for each $j$, the coefficients,
$\gamma_{1j},\ldots,\gamma_{Nj}$, are nonnegative and sum to 1.
The functional $f$ is a known, problem-dependent (possibly
nonlinear) function of the linear combination of the components $\X$
and some unknown parameters,
$\theta_j$. Each $\bolds{\varepsilon}_j$ is a vector of random errors.
The set of target parameters for each observed data vector, $\Y_j$, is
$\{\rho_j, \theta_i\}$, where $\rho_j = \sum_{i=1}^N \gamma
_{ij}\pi_i$ is a function of the model weights, $\gamma$, and
intrinsic parameters, $\pi$, of the theoretical components.

For large $N$, it is impossible to use model (\ref{eqngenmodel}) to
estimate each $\{\rho_j,\theta_j\}$ due to the large computational
cost. Our goal is to find a set of prototypes $\bolds{\Psi} =
[\bolds{\Psi}_1,\ldots,\bolds{\Psi}_K]$, where $K \ll N$, that can
accurately estimate the target parameters $\{\rho_j,\theta_j\}$ for each
observed $\Y_j$, using the model
%
\begin{equation}
\label{eqnmodel}
\Y_{j} = f\Biggl(\sum_{k=1}^K \beta_{kj} \bolds{\Psi}_{k} ;
\bolds{\theta}_{j} \Biggr),
\end{equation}
where $\beta_{1j},\ldots,\beta_{Kj}$ are nonnegative component weights
such that $\sum_k\beta_{kj} = 1$ for all $j$. Naturally, our estimate
of $\rho_j$ is
%
\begin{equation}
\label{eqnparamest}
\widehat{\rho}_j = \sum_{k=1}^K \widehat{\beta}_{kj}\sum_{i=1}^N
\alpha_{ik}\pi_i,
\end{equation}
where the $\widehat{\beta}_{jk}$ are estimated using the model (\ref
{eqnmodel}), and $\bolds{\alpha}$ is an $N$ by $K$ matrix of
nonnegative coefficients that defines the prototypes from the
dictionary of components by
%
\begin{equation}
\label{eqnproto}
\bolds{\Psi} = \X\bolds{\alpha}.\vadjust{\goodbreak}
\end{equation}
The coefficients $\bolds{\alpha}$ are constrained such that each of
the prototypes, $\bolds{\Psi}_k$, resides in a region of the
theoretical component space, $R_k \in\mathcal{X}$, with nonzero
probability, $P_X(R_k) > 0$, over all plausible values of the physical
parameters used to generate~$\X$. This constraint is enforced to
ensure the physical plausibility of the prototypes, $\bolds{\Psi}$,
and their parameters. 
If our prototype basis were to include components that are disallowed
by the physical models that generated $\X$, then the parameter
estimates for the observed data would be uninterpretable.\vspace*{-1pt} 

\section{Methods for prototyping}
\label{secmeth}

The usual method used to choose a basis for estimating target
parameters from the signal model is to select prototypes from a regular
grid in the physical parameter space. Examples of such bases are those
found in \citet{CF2005} and \citet{asa2007}, both of whom
employ SSPs on regular grids of age and metallicity to estimate SFH
parameters. In this section we propose methods that use the set of
physical components, $\X$, to construct a prototype basis in a
principled manner. In Section \ref{secsim} we compare the proposed
basis selection methods via simulations, and show that regular
parameter grids tend to yield suboptimal parameter estimates.\vspace*{-1pt}


\subsection{Quantization of model space}
\label{ssquant}

$\!\!\!$For problems of interest, practical \mbox{fitting}
of theoretical models to noisy data requires a finite set of
prototypes. The question
becomes how to best choose this set of prototypes, that is,~how to
\textit{quantize the
model space}. Here, instead of quantizing the parameter space~by
choosing uniform parameter grids, we propose methods that quantize the
vector space $\mathcal{X}$ of theoretical model-produced data. The
idea behind this~ap\-proach is that under the presence of noise,
components with similar functio\-nal forms will be indistinguishable, so
that it is better to choose prototypes that are approximately evenly
spaced in $\mathcal{X}$ (rather than evenly spaced in the parameter
space). By replacing the theoretical models in each neighborhood by
their local average, the model quantization approach is optimal for
treating degeneracies because it allows a slight increase in bias to
achieve a large decrease in variance of the target parameter estimates.
The increase in estimator bias should be small because more prototypes
are included in parameter regions where we can better discern the
theoretical data curves of the components, allowing for precise
parameter estimates in those regions and coarser average estimates in
degenerate regions. If, instead, multiple components in our dictionary
were to have very similar theoretical data curves but different
parameter values, then, in the absence of any other method of
regularization, we would have difficulty breaking the degeneracy no
matter how many prototypes we include in that region of the parameter
space, causing increased parameter estimator variance and higher
statistical risk.\looseness=-1\vspace*{-1pt}

\subsubsection{$K$-means and diffusion $K$-means}
The basic idea here is to quantize the vector space\vadjust{\goodbreak} or support of
model-produced data with respect to an appropriate metric and prior
distribution. The vector quantization approach can be formalized as follows:

Suppose that $\mathbf{X}_1, \ldots, \mathbf{X}_N$ is a sample from
some distribution $P_X$ with support $\mathcal{X} \subset\mathbb
{R}^p$. The support $\mathcal{X}$ often has some lower-dimensional
structure, which we refer to as the lower-dimensional \textit{geometry} of
$\mathcal{X}$. 
Fix an integer $K<N$. To any dictionary
$A=\{\mathbf{a}_1,\ldots, \mathbf{a}_K\}$ of prototypes, we can
assign a~cost
%
\begin{equation} \label{eqKmeanscost}
W(A,P_X)=\int{\min_{\mathbf{a}\in A} }\|\mathbf{x} - \mathbf{a}\|^2
P_X(d\mathbf{x}).
\end{equation}
%
Let $\mathcal{B}_k$ denote all sets of
the form $B=\{\mathbf{b}_1,\ldots, \mathbf{b}_K\}$ with $\mathbf
{b}_j\in\mathbb{R}^p$.
Define the optimal dictionary of $K$ prototypes as the cluster centers
\[
\bolds{\Psi} = \mathop{\arg\min}_{B\in\mathcal{B}_k} W(B,P_X).
\]
In practice, we estimate $\bolds{\Psi}$ from model-produced data
$\mathbf{X}_1, \ldots, \mathbf{X}_N$ according to
\[
\widehat{\bolds{\Psi}} = \mathop{\arg\min}_B W(B,\widehat{P}_X),
\]
where $\widehat{P}_X$ is the empirical distribution. This estimate is
found by Lloyd's $K$-means (KM) algorithm. To simplify the notation, we
will henceforth skip the hat symbol on all estimates.

The empirical $K$-means solution corresponds to allocating each
$\mathbf{X}_i$ into subsets $S_1,\ldots,S_K$, where the $K$ centroids
define the prototypes.
In the definition of the prototypes in (\ref{eqnproto}), this reduces to
%
\begin{equation}
\label{eqnkm}
\alpha_{ik} = \cases{
\dfrac{1}{|S_k|}, &\quad if $i \in S_k$, \vspace*{2pt}\cr
0, &\quad else.}
\end{equation}
%
Potential problems to this approach are the following: (1) the KM
prototypes will adhere to the design density on $\mathcal{X}$, and (2)
for small $K$, estimated prototypes could fall in areas that $P_X$
assigns probability zero. The first issue can be corrected using a
weighted $K$-means approach or a method such as uniform subset
selection (Section \ref{sssuss}). However, often the density on~$\mathcal{X}$
corresponds to a prior distribution on the physical
parameters, meaning it is often desirable to adhere to its design density.
To remedy the latter issue, we could select as prototypes the $K$ data
points that are closest to each of the centroids. We see in simulations
that this approach tends to yield slightly worse parameter estimates
than the original $K$-means formulation. We attribute this to the
smoother sampling of parameter space achieved by the original KM
formulation, which averages the parameters of components with similar
theoretical data, effectively decreasing the variability of the
parameter estimates.

If the theoretical data are high dimensional, we might choose to first
learn the low-dimensional structure of $\X$ and then employ $K$-means
in this reduced space. This would permit us to avoid quantizing
high-dimensional data, where $K$-means can be problematic due to the
curse of dimensionality. This failure occurs because the theoretical
data are extremely sparse in high dimensions, causing the distances
between similar components to approach the distances between unrelated
objects. To remedy this, we suggest the use of the diffusion map method
for nonlinear dimensionality reduction
[\citet{coif2006}, \citet{lafo2006}]. In other words, we
transform the model data into a lower-dimensional representation where
we apply $K$-means (diffusion $K$-means, DKM).
Formally, this corresponds to substituting (\ref{eqKmeanscost}) with
the cost function
%
\begin{equation}
W(\phi, A,P_X)=\int{\min_{\mathbf{a}\in A}} \|\phi(\mathbf{x})-\phi
(\mathbf{a})\|^2 P_X(d\mathbf{x}),
\end{equation}
where $\phi$ is a data transformation defined by diffusion
maps.\footnote{Software for diffusion maps and diffusion $K$-means is
available in the \texttt{diffusionMap~R} package, which can be downloaded
from
\texttt{\href{http://cran.r-project.org/web/packages/diffusionMap/index.html}{http://cran.r-project.org/web/packages/}
\href{http://cran.r-project.org/web/packages/diffusionMap/index.html}{diffusionMap/index.html}}.}

\subsubsection{Uniform subset selection}
\label{sssuss}

In the theoretical model data quantization approach the goal is to have
prototypes regularly spaced in~$\mathcal{X}$, where~$\mathcal{X}$ is
the support of~$P_X$. With this heuristic in mind, we devise the
uniform subset selection (USS) method, which sequentially chooses the
component $\X_i \in\X$ that is furthest away from the closest
component that has already been chosen.
Because the choice of distance metric is flexible, USS can be tailored
to deal with many data types and high-dimensional data.
Unlike $K$-means, USS is not influenced by differences in the density
of components across~$\mathcal{X}$. However, USS typically chooses
extreme components as prototypes because in each successive selection
it picks the furthest theoretical data curve from the active set. In
simulations, USS produces poor parameter estimates due to its tendency
to select extreme components.
%

\subsection{Sparse coding approaches}
Most standard sparse coding techniques do not apply for the prototyping
problem. Without the appropriate constraints, the prototype basis
elements will be nonphysical and the subsequent parameter estimates
will be nonsensical (see Section \ref{sssother}). There are methods
related to sparse coding that enforce the proper constraints to ensure
that prototype basis elements reside within the native data space (see
Sections \ref{secaa} and \ref{ssssss}), but these generally do not perform
well for target parameter estimation because their objective of optimal
data reconstruction---and not estimation of the target
parameters---forces these methods to choose extreme prototypes.

\subsubsection{Archetypal analysis}
\label{secaa}

Archetypal analysis (AA) was introduced by Cutler and Breiman
(\citeyear{cutl1994}) as a
method of representing each data point as a~linear mixture of
archetypal examples, which themselves are linear mixtures of the
original component dictionary. The method searches for the set of
archetypes $\bolds{\Psi}_1,\ldots,\bolds{\Psi}_K$ that satisfy (\ref
{eqnproto}) and minimize the residual sum of squares (RSS)
%
\begin{eqnarray}
\label{eqnaa}
\mathrm{RSS} &=& \sum_{i=1}^N \Biggl\| \X_i - \sum_{k=1}^K \beta_{ik}
\bolds{\Psi}_k\Biggr\|^2\\
&=& \sum_{i=1}^N \Biggl\| \X_i - \sum_{k=1}^K \beta_{ik}\sum_{j=1}^N
\alpha_{jk}\X_j\Biggr\|^2,
\end{eqnarray}
where $\sum_{k=1}^K \beta_{ik} = 1$ for all $i$ and $\beta_{ik}\ge
0$ for all $i$ and $k$. To minimize the RSS criterion, an alternating
nonnegative least squares algorithm is employed, alternating between
finding the best $\bolds{\beta}$'s for a set of prototypes and
finding the best prototypes ($\bolds{\alpha}$'s) for a set of
$\bolds{\beta}$'s.
This computation scales linearly in the number of dimensions of the
original theoretical data, with computational complexity becoming
prohibitive for dimensionality more than 500
[\citet{ston2002}]. 

Once there are as many prototypes, $K$, as the number of data points
that define the boundary of the convex hull, any element in the
dictionary can be fit perfectly with a linear mixture of the
prototypes, yielding a RSS of 0. If we try to pick more prototypes than
the number of data points that define the boundary of the convex hull,
then the AA algorithm will fail to converge because $\beta$ becomes
noninvertible, preventing the iterative algorithm to find the optimal
set of prototypes, $\bolds{\Psi} = \beta^{-1}\X$, given the
current $\beta$. We have experimented with using the Moore--Penrose
pseudoinverse to perform this operation, but it is usually ill-behaved
when $\beta$ is noninvertible. This upper bound on the number of AA
prototypes is a serious drawback to using AA as a prototyping method
because often the complicated nature of the data generating processes
necessitates the use of larger prototype bases.

Prototypes found by AA are optimal in the sense that they minimize the
RSS for fitting noiseless, linear mixtures of the $\X$'s. This is the
case because~AA prototypes are found along the boundary of the convex
hull formed by the $\X$'s [see \citet{cutl1994}]. Unlike AA, our
objective is not\vadjust{\goodbreak} to minimize RSS, but to minimize the error in the
derived parameter estimates.
Archetypal analysis achieves suboptimal results in the estimation of
$\bolds{\rho}$ because it only samples prototypes from the boundary
of the component space,~$\mathcal{X}$, focusing attention on extreme
cases while
disregarding large regions of~$\mathcal{X}$.
In Section \ref{secsim} we show using simulated data that AA is
outperformed by the model quantization approach for estimating the
target parameters from the signal model parameters.


\subsubsection{Sparse subset selection}
\label{ssssss}
We introduce the method of sparse subset selection (SSS), whose goal is
to find a
subset of the original dictionary, $\bolds{\Psi} \subset\X$, that
can reconstruct
$\X$ in a linear mixture setting. This method is motivated by sparse
coding in that it seeks the basis that
minimizes a~regularized reconstruction of $\X$, where the
regularization is chosen to select a~subset of the columns of $\X$.

Recently, \citet{oboz2009} introduced a method of variable
selection in a high-dimensional multivariate
linear regression setting. Their method uses a penalty on the
$\ell_1/\ell_q$ norm, for $q>1$, of the matrix of regression
coefficients in such
a way that induces sparsity in the rows of the coefficient matrix. We
can, in a straightforward way, adapt their method to select a~subset
of columns of $\X$ to be used as prototypes. Our objective function~is
%
\begin{equation}
\label{eqnSSS1}
\mathop{\arg\min}_{\B} \biggl\{ \frac{1}{2N} \| \X- \X\B\|_F^2 + \lambda_k
\|\B\|_{\ell_1/\ell_q} \biggr\},
\end{equation}
where \mbox{$\|\cdot\|_F$} is the Frobenius norm of a matrix, and the $\ell
_1/\ell_q$ penalty is defined as
%
\begin{equation}
\label{eqnSSS2}
\|\B\|_{\ell_1/\ell_q} = \sum_{i=1}^N \Biggl(\sum_{j=1}^N
b_{ij}^q\Biggr) ^{1/q} = \sum_{i=1}^N \|b_i\|_q
\end{equation}
so that sparsity is induced in the \textit{rows} of $\B$, the $N$ by
$N$ matrix of nonnegative mixture coefficients. Additionally, $\B$ is
normalized to sum to 1 across columns.
The basis, $\bolds{\Psi}$, is defined as the columns of $\X$ that
correspond to nonzero rows of $\B$ ($\bolds{\alpha}$ is the
corresponding indicator variable). The
parameter $\lambda_k$ controls the number of prototypes in our SSS set
$\bolds{\Psi}$.

To perform the optimization (\ref{eqnSSS1}), we use the \texttt{CVX
Matlab} package [\citet{cvx}]. Setting $q=2$, we recast the
problem as a second-order cone problem with the additional constraints
of nonnegativity and column normalization of $\B$ [see
\citet{boyd2004}]. The current implementation cannot solve
problems for large $N$. In Section \ref{sscomp} we show, for a~small
problem, that SSS has behavior similar to archetypal analysis in that
it selects prototypes from the boundary of the convex hull of $\X$.
Like~AA, SSS is not a good method for target parameter
estimation.

\subsubsection{Some methods not useful for prototyping}
\label{sssother}
There are other methods for sparse data representation that fail to
work for prototype selection. These methods are not applicable to this
problem because they do not select prototypes that reside in regions of
$\mathcal{X}$ with nonzero probability $P_X$. The failure to obey this
constraint means that the chosen prototypes in general will not be
\textit{physical}, meaning that either their theoretical data or
intrinsic parameters are disallowed. For instance, in the SFH problem,
this could lead us to use prototypes whose spectra have negative photon
fluxes or whose ages are either negative or greater than the age of the
Universe. Using such uninterpretable prototypes to model observed data
produces parameter estimates that are nonsensical.

We mention two popular methods for estimating small bases from large
dictionaries, $\X$, and describe why they are not useful for prototyping:

In \textit{standard sparse coding} [\citet{olsh1996}], the goal is
to find a~decomposition of the matrix $\X$, in which the hidden
components are sparse. Sparse coding combines the goal of small
reconstruction error along with sparseness, via minimization of
%
\begin{equation}
\label{eqnsc}
C(\bolds{\Psi},\A) = \frac{1}{2}\|\X- \bolds{\Psi} \A\|^2 +
\lambda\sum_{ij}|a_{ij}|,
\end{equation}
where the trade-off between $\ell_1$ sparsity in the mixture
coefficients $\A$, and accurate reconstruction of $\X$, is controlled
by $\lambda$. However, there are no constraints on the sign of the
entries of $\A$ or $\bolds{\Psi}$, meaning that prototypes with
nonphysical attributes are allowed.

\textit{Nonnegative Matrix Factorization} (\textit{NMF})
[\citet{lee2001}, \citet{paat1994}] is a related technique
that includes
strict nonnegativity constraints on all coefficients $a_{ij}$ and
$\Psi_{jk}$ while minimizing the reconstruction of $\X$,
%
\begin{equation}
\label{eqnnmf}
\mathop{\arg\min}_{\Psi, \A} \biggl\{ \frac{1}{2}\|\X- \bolds{\Psi} \A
\|^2 \biggr\}.
\end{equation}
This construction is different than our prototype definition in (\ref
{eqnproto}), where $\bolds{\Psi} = \X\bolds{\alpha}$. To
reconcile the two, we see that, since $N>K$, $\bolds{\alpha}$ is the
right inverse of $\A$:
%
\begin{equation}
\bolds{\alpha} = \A(\A^T\A)^{-1},
\end{equation}
which exists if $\A$ is full rank. However, under this formulation,
the $\alpha_{ij}$ are not constrained to be nonnegative and the
resultant prototypes are not constrained to reside in
$\mathcal{X}$. Thus, NMF is not useful for prototyping. Note that
archetypal analysis avoids this problem by enforcing the further
constraint that the prototypes be constrained linear combinations of
$\X$.

%

\subsection{Comparison of prototypes}
\label{sscomp}

\begin{figure}

\includegraphics{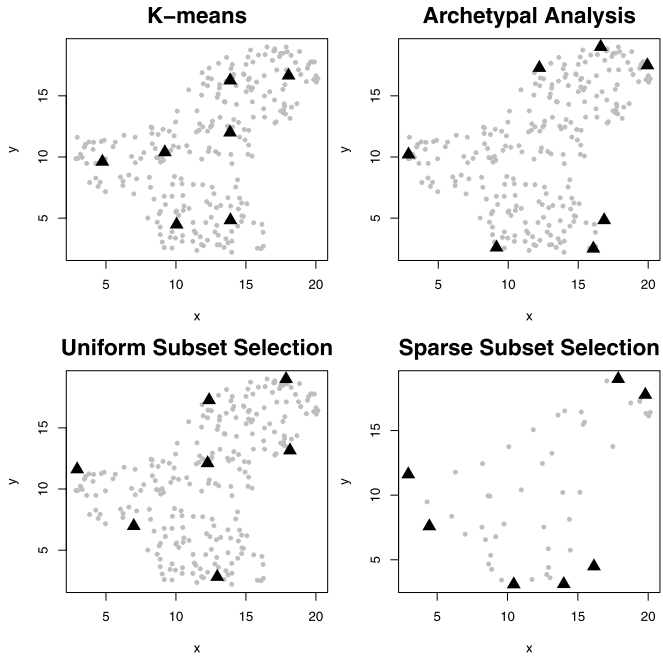}

\caption{Distribution of prototypes (red $\blacktriangle$'s) for four
different methods when applied to the 250 theoretical data objects in
the \texttt{toy} data set (grey $\bullet$'s). $K$-means evenly samples
the native data space while the other methods focus more attention to
the boundary of the space.} \label{figtoyproto}
\end{figure}

We apply four prototyping methods to the two-dimensional data set
\texttt{toy} in the \texttt{archetypes R}\vadjust{\goodbreak} package.\footnote{Available from
\texttt{CRAN} at \url{http://cran.r-project.org/web/packages/archetypes}.} We
treat each 2-D data point, $\X_i$, as model-produced theoretical data.
Plots of this dictionary of data and the selected prototypes for four
different prototyping methods, using $K=7$, are in Figure \ref
{figtoyproto}. $K$-means places prototypes evenly spaced within the
convex hull of the data. USS also evenly allocates the prototypes, but
%
places many along the boundary of the native space. Archetypal analysis
and SSS place all prototypes on the boundary of the convex hull. Note
that for more than 7 prototypes, the archetypal analysis algorithm does
not converge to a solution.

\section{Simulated examples}
\label{secsim}

In this section we test the effectiveness of the prototyping methods
for estimating a set of target parameters using simulated data. The
first test set is the toy example of zero-mean Gaussian curves
discussed in Section~\ref{secex}. The second simulation experiment is
a set of realistic galaxy spectra created to mimic the SDSS data that
we later analyze in Section \ref{secsdss}.

\subsection{Gaussian curves}
\label{secnorm}

We begin with the example introduced in Section~\ref{secex}. We
simulate a database of $N=157, \mu=0$ Gaussian curves, $\X_1,\ldots,\allowbreak\X
_N$, on a fine grid of $\sigma=(\sigma_1,\ldots,\sigma_N)$ from 0.2 to
8 in steps of 0.05 (see Figure~\ref{fignormcurv}). Each $\X_i$ is
represented as a vector of length 321. From this database, we simulate
a set of 100 data vectors, $\Y_1,\ldots,\Y_{100}$, from the model
%
\begin{equation}
\label{eqnnorm}
\Y_{j} = \sum_{i=1}^N \gamma_{ij} \X_{i} + \varepsilon_{j},
\end{equation}
where the mixture coefficients, $\gamma_{ij}\ge0$, sum to unity for
each $j$ and have at most 5 nonzero entries for each $j$. The noise
vectors, $\varepsilon_{j}$, are i.i.d. normal zero-mean with standard
deviation 0.05.

\begin{figure}

\includegraphics{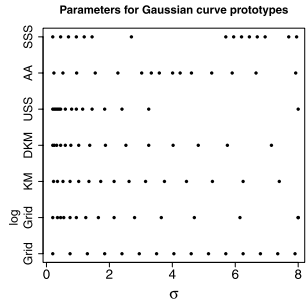}

\caption{Distribution of $K=15$ prototype $\sigma$ values for seven
different prototyping methods applied to the Gaussian curves example.
The methods are the following: Grid-regular~$\sigma$ grid, log
Grid-regular $\log(\sigma)$ grid, KM---$K$-means, DKM---diffusion $K$-means,
USS---uniform subset selection, AA---archetypal analysis, and SSS---sparse
subset selection.}
\label{fignormsig}
\end{figure}

From $\X_1,\ldots,\X_N$, we generate bases of prototypes using six
different methods described in Section \ref{secmeth}. To explore the
differences in each of these methods, we plot (Figure \ref{fignormsig})
the distribution of $K=15$ prototype $\sigma$ values. The model
quantization methods (KM, DKM, USS) find more prototypes with small
$\sigma$ values. The AA and SSS methods place more prototypes at the
extreme values of $\sigma$ (note that for~SSS, we ran the algorithm on
a coarser grid of 32 Gaussian curves).

\begin{figure}

\includegraphics{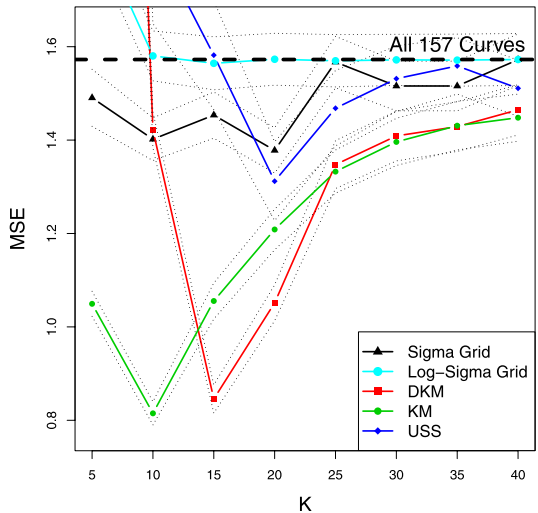}

\caption{MSE for the estimation of $\sigma$ for the Gaussian curve
example. Plotted is the MSE for using a regular parameter grid,
$K$-means (KM), diffusion $K$-means (DKM) and archetypal analysis (AA)
prototype bases. Both DKM and KM achieve significantly better~$\bar{\sigma}$
estimates than a regular parameter grid and outperform
estimates obtained by using all 157 Gaussian curves in the original
dictionary. For each $K$, the MSE is averaged across 25 repetitions of
the experiment. Point-wise 68\% confidence bands are shown as dotted
lines.} \label{figsigmse}
\end{figure}

To evaluate each of the methods, we compare their ability to estimate
the average $\sigma$ for each $\Y_j$, defined as
%
\begin{equation}
\bar{\sigma}_j = \sum_{i=1}^N \gamma_{ij} \sigma_i.
\end{equation}
For each choice of basis, we fit the observed data using nonnegative
least squares.\footnote{We use the \texttt{nnls R} package, which uses the
Lawson--Hanson nonnegative least squares implementation [\citet
{lawson1995}].}
In Figure \ref{figsigmse} the MSE for $\bar{\sigma}$ estimation for
$K$-means, diffusion $K$-means, USS and uniform $\sigma$-grid and
$\log(\sigma)$ grid bases is plotted as a function of $K$. SSS is not
plotted because it yields parameter estimates with MSE $> 2$. AA is not
plotted because it only converges for $K \le15$, and performs worse
than the $\sigma$ grid for those values. KM and DKM outperform the
regular parameter grids, USS, and AA prototype bases. KM achieves a
minimum MSE, averaged over 25 trials, of 0.815 at $K=10$ prototypes.
DKM achieves a minimum MSE of 0.846 at \mbox{$K=15$} prototypes, while the
uniform $\sigma$ grid achieves\vadjust{\goodbreak} a minimum MSE of 1.378, 1.7 times
higher than the best MSE for KM. Results for AA and SSS are not plotted
because AA only converges for $K \le15$ prototypes, and SSS is too
computationally intensive to run on the entire dictionary of curves; at
$K=15$, neither method outperforms a~uniform $\sigma$ grid.

An interesting observation in Figure \ref{figsigmse} is that the
minimum MSE for estimating ${\bar{\sigma}}$ is achieved for $K=10$ KM
prototypes. As the number of prototypes increases from 10, the KM $\bar
{\sigma}$ estimates worsen. This exemplifies the bias-variance
trade-off in the estimation procedure: for $K > 10$, the increased
variance of the estimates is larger than the reduction in squared-bias.
Estimates of $\bar{\sigma}$ from four of the five prototype bases
plotted in Figure~\ref{figsigmse} outperform the estimates found by
fitting each $\Y_j$ as a mixture of all 157 original component curves.
Over the 25 repetitions of the simulations, the~$\gamma_{ij}$ which
are positive,
that is, the $\mathbf{X}_i$ that receive any weight, vary widely. These results
demonstrate that a single, judiciously chosen, reduced basis can reproduce
a wide range of truths and return accurate parameter estimates
with reduced variance.

\subsection{Simulated galaxy spectra}

We further test the performance of each prototyping method using
realistic simulated galaxy spectra. Starting with a database, $\X$, of
1,182 SSPs from the models of \citet{BC2003} (see Section
\ref{secsfh}), we generate simulated galaxy spectra using the
model~(\ref{cfmodel}). The SSPs are generated from 6 different metallicities
and a fine sampling of 197 ages from 0 to 14 Gyrs. We use a
prescription similar to \citet{chen2009} to choose the physical
parameters of the simulations, altered to have higher contribution from
younger SSPs. The basic physical components of the simulation are as
follows:
\begin{longlist}[(1)]
\item[(1)] A star formation history with exponentially
decaying star formation rate (SFR): SFR $\propto\exp(\gamma t)$.
Here, $\gamma>0$, so the SFR is exponentially declining with time, as
$t$ is the age of
the SSP today.
\item[(2)] We allow
$\gamma$ to vary between galaxies. For each galaxy we draw $\gamma$
from a uniform distribution between 0.25 and 1 $\mathrm{Gyr}^{-1}$.
\item[(3)] The time $t_{\mathrm{form}}$ when a galaxy begins star formation is
distributed uniformly between 0 and 5.7 Gyr after the Big Bang,
where the Universe is assumed to be 13.7 Gyr old.
\item[(4)] We allow for starbursts, epochs of increased SFR, with equal
probability at all times.
The probability a starburst begins at time $t$ is constructed so that
the probability of no starbursts in the life of the galaxy is 33\%.
The length of each burst is distributed uniformly between 0.03 and
0.3 Gyr and the fraction of total stellar mass formed in the burst
in the past 0.5 Gyr is distributed log-uniformly between 0 and 0.5.
The SFR of each starburst is constant throughout the length of the burst.
\end{longlist}
Each galaxy spectrum is generated as a mixture of SSPs of up
to 197 time bins, with a uniformly drawn metallicity in each bin. We
draw the reddening parameter ($A_V$) and velocity dispersion ($\sigma
_0$) from empirical distributions over a plausible range of each parameter.
We simulate\vspace*{1pt} 100 galaxy spectra with i.i.d. zero-mean Gaussian noise
with $\mbox{S}/\mbox{N}=10$ at $\lambda_0=4020$ \AA.

\begin{figure}

\includegraphics{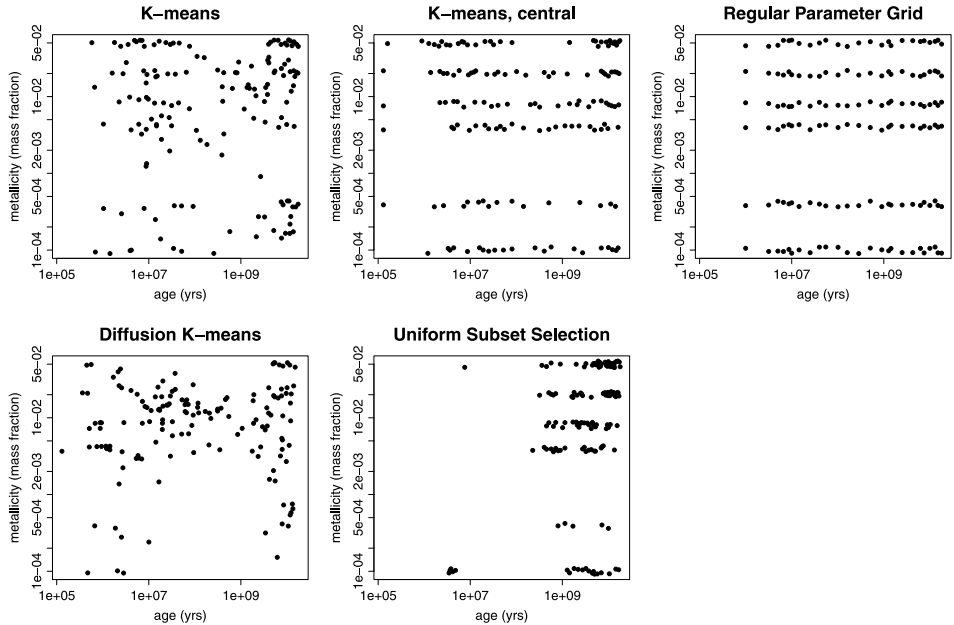}

\caption{Distribution of $(t,Z)$ of several prototype bases of SSPs,
$K=150$. All bases were derived using a database of 1,182
model-produced SSPs. Each of the methods more heavily samples
prototypes with large age and large metallicity.} \label{figssppar150}
\end{figure}

We apply the methods in Section \ref{secmeth} to choose SSP prototype
bases from~$\X$. In Figure \ref{figssppar150} the distributions of
the SSP prototype ages and metallicities for $K=150$ prototype bases
are plotted along with the regular parameter grid used by
\citet{asa2007}. Each method highly samples the older, higher
metallicity SSPs and typically only includes a few prototypes with low
age and low metallicity. This is reasonable because older, higher
metallic SSP spectra change more with respect to changes in age and
metallicity. Any method for prototyping based on the model-produced
data will detect this difference and sample these regions of the
parameter space more highly.\looseness=1

\begin{figure}

\includegraphics{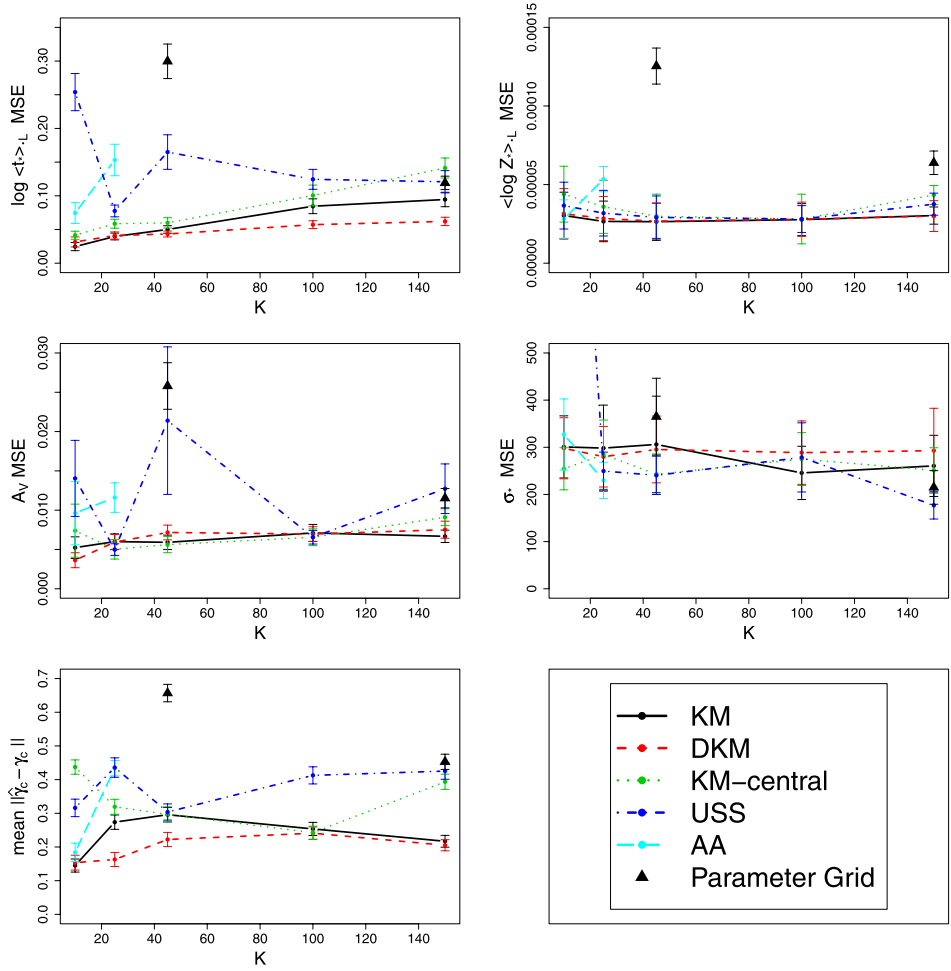}

\caption{Errors in physical parameter estimates for galaxy simulations
using prototype techniques: $K$-means (KM), diffusion $K$-means (DKM),
centroid $K$-means (KM-central), USS, AA, and a regular parameter grid.
MSEs are plotted for bases of size $K=10$, 25, 45, 100 and 150. The
regular parameter grids are from
Cid Fernandes et al. (\protect\citeyear{CF2005}) ($K=45$) and
Asari et al. (\protect\citeyear{asa2007}) ($K=150$). Each prototyping method finds more accurate
SFH parameter estimates than the two regular parameter grids.}
\label{figwild2}
\end{figure}

Each simulated galaxy spectrum is fit using the \texttt{STARLIGHT}
software with each prototype basis. To assess the performance of each
method, we compare the accuracy of their parameter estimates. In Figure
\ref{figwild2} we plot the MSE of the estimates of $\log\langle t_*
\rangle_L, \langle\log Z_*\rangle_L, A_V$ and $\sigma_*$ and the
average error of the coarse-grained population vector estimate,
%
$\widehat{\gamma}_c$, measured by the average~$\ell_2$ distance to
the true $\gamma_c$.
Each prototype method outperforms the regular parameter grid prototype
bases, often by large margins, especially for $K=45$. Between the
different prototyping methods there does not appear to be a clear
winner, though diffusion $K$-means bases achieve the lowest or
second-lowest MSE for 4 of the 5 parameters. $K$-means also achieves
accurate estimates for each of the parameters, and always beats or ties
the $K$-means-central estimates. Both USS and AA yield inaccurate
estimates for all parameters except $\langle\log Z_*\rangle_L$ and
$\sigma_*$. SSS could not be run on such a large dictionary of SSPs.
Overall, small bases achieve better estimates of $\log\langle t_*
\rangle_L, A_V$ and $\gamma_{\mathrm{c}}$, but this likely will not be the
case for real galaxies, whose SFHs are more complicated and diverse
than the simulation prescription used.


\section{Analysis of SDSS galaxies}
\label{secsdss}

\begin{figure}

\includegraphics{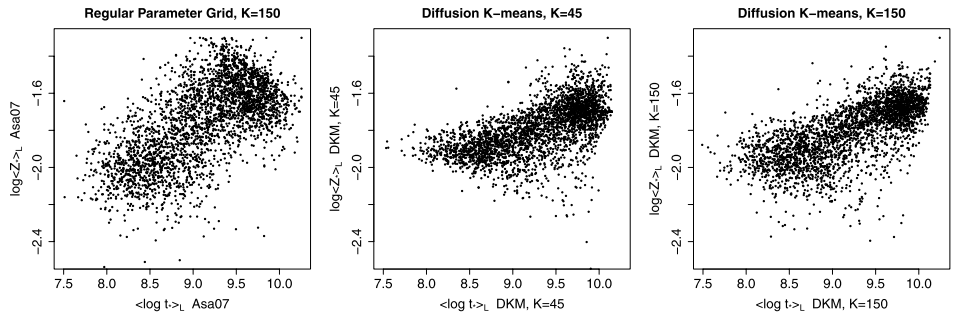}

\caption{Estimates of $\log\langle t_* \rangle_L$ versus $\langle
\log
Z_*\rangle_L$ for a set of 3046 galaxies observed by the SDSS,
estimated using \texttt{STARLIGHT} with three different prototype bases.
From left to right, bases are as follows: regular parameter grid from
Asari et al. (\protect\citeyear{asa2007}) with $K=150$, diffusion $K$-means $K=45$, and
diffusion $K$-means $K=150$. Estimates from diffusion $K$-means bases
show much less spread in the direction of the well-known
age-metallicity degeneracy in galaxy population synthesis studies.}
\label{figsdss}
\end{figure}

Prototyping methods are used to estimate the SFH parameters from the
SDSS spectra of a set of 3046 galaxies in SDSS Data Release 6
[\citet{adel2008}]. For more detailed information about the data
and preprocessing steps, see \citet{rich2009}. In Figure
\ref{figsdss} we plot the estimated $\log\langle t_* \rangle_L$ versus
$\langle\log Z_*\rangle_L$ for each galaxy using three basis choices:
the regular parameter grid of \citet{asa2007} (Asa07, $K=150$),
DKM with $K=45$, and DKM with $K=150$.\vfill\eject

There are several differences in the estimated $\langle\log Z_*\rangle
_L-\log\langle t_* \rangle_L$ relation for each basis. First, both
diffusion $K$-means bases produce estimates that are tightly spread
around an increasing trend while the Asa07 estimates are more diffusely
spread around such a trend. The direction of discrepancy in the Asa07
estimates from the trend corresponds exactly with the direction of a
well-known spectral degeneracy between old, metal-poor and young,
metal-rich galaxies
[\citet{wort1994}]. This suggests that the observed variability
along this direction is not due to the physics of these galaxies, but
rather is caused by confusion stemming from the choice of basis
[in \citet{rich2009} we verified that diffusion $K$-means SFH
estimates have a~decreased age-metallicity degeneracy, using simulated
galaxy spectra]. Second, the $K=45$ diffusion $K$-means basis estimates
no young, metal-poor galaxies, whereas the other bases do. This
suggests that this small number of prototypes is not sufficient to
cover the parameter space; particularly, young, metal-poor SSPs have
been neglected in the $K=45$ diffusion $K$-means basis. Finally, the
overall trend between $\log\langle t_* \rangle_L$ versus $\langle
\log Z_*\rangle_L$ differs substantially between the regular grid and
diffusion $K$-means basis, suggesting that SFH parameter estimates are
sensitive to the choice of basis and that downstream cosmological
inferences will depend heavily on the basis used. 

Recently, we have estimated the SFH parameters for all 781,692 galaxies
in the SDSS DR7 [\citet{abaz2009}] main sample or LRG sample. This
subset of DR7 galaxies was chosen for analysis because it was targeted
for spectroscopic observation, and thus has a well defined selection
function [\citet{stra2002}]. We estimated the parameters using
\texttt{STARLIGHT} with a diffusion $K$-means basis of size $K=150$. The
computational routines took nearly 5 CPU years to analyze the entire
data set, which includes preprocessing of the data, estimating the SFH
parameters for each, and compiling the catalog of estimates. The
computations were performed in parallel on the 1,000-core
high-performance FLUX cluster at the University of Michigan. Results
of this analysis are in preparation [\citet{rich2010}] and will be
published shortly. These SFH estimates will be used to constrain
cosmological models that concern the formation and evolution of
galaxies and the history and fate of the Universe.

There is also ongoing work into approaches to quantifying the statistical
uncertainty in the resulting parameter estimates. This is a critical, but
challenging, component. The basic approach to be employed will exploit the
massive amount of data by inspecting the amount of variability in parameter
estimates in small neighborhoods in the space of galaxy spectra. An
additional regression model will be fit, with the parameter estimates as
the response, and the spectrum as the predictor. In previous work
[\citet{rich2009a} and \citet{free2009}],
we have fit models of exactly this type, using
galaxy spectra or colors to predict redshift. As was the case in that
work, we will
smooth the parameter estimates in the high-dimensional space
to obtain an estimator with lower variance. Equally important, this will
yield a natural way of estimating the uncertainty in the estimator, by
inspecting the variance of the residuals of the regression fit.

\section{Conclusions}
\label{secconc}

We have introduced a prototyping approach for the common class of
parameter estimation problems where observed data are produced as a
constrained linear combination of theoretical model-produced
components, and the target parameters are derived from the parameters
in the signal model. The usual approach to this type of problem is to
use models on a regular grid in parameter space. In this paper we have
introduced approaches that use the properties of the theoretical data
from the dictionary of components to estimate prototype bases. These
approaches include: quantizing the component model data space using
$K$-means, selecting prototypes uniformly over the space of theoretical
component data, and estimating prototype bases that minimize the
reconstruction error of the components.

Our main findings are the following:
\begin{itemize}
\item The quantization methods presented in this paper achieve better
parameter estimates than the approach of using prototypes from a
regular parameter grid, as shown in multiple simulations. The regularization
that results from a reduced basis leads to reduced variance in the
parameter estimates, without sacrificing accuracy. This is the case
because components with similar theoretical data will be indiscernible
under the presence of noise, making it crucial that prototypes be
spread out evenly in theoretical data space, inducing a large decrease
in variance of the target parameter estimates. If bases are too small,
then the parameter estimates suffer from large bias because important
regions of model space are neglected.
\item Standard sparse coding methods are not appropriate for this class
of problem. Without the proper constraints, these methods do not find
prototypes that are physically-plausible. Even with these constraints,
these methods select prototypes around the boundary of the data
distribution, which is good for data reconstruction but not for target
parameter estimation.
\item For a complicated problem in astrophysics---estimating the
history of star formation for each galaxy in a large database---we
obtain more accurate parameters (in simulations) using the model
quantization approach than using regular parameter grids. When applied
to the real data, these different prototyping approaches produce
markedly different results, showing the importance of prototype basis selection.
\end{itemize}

%
%
%
%

%
%
%
%



\printaddresses

\end{document}